\documentclass[journal]{IEEEtran}
\IEEEoverridecommandlockouts
\usepackage{lmodern,textcomp}
\usepackage{graphics}
\usepackage{multicol}
\usepackage{lipsum}
\usepackage{color}
\usepackage{amsthm}
\usepackage{amssymb}
\usepackage{mathtools}
\usepackage{amsbsy}
\usepackage{setspace}
\usepackage{float}
\usepackage{lettrine}
\usepackage{newclude}
\usepackage[normalem]{ulem}
\usepackage{latexsym}
\usepackage{multirow}
\usepackage{rotating}
\usepackage{booktabs}
\usepackage{wasysym}
\usepackage{mathrsfs}
\usepackage{bbm} 
\usepackage{epsfig,cite,amsfonts,psfrag}
\usepackage{apptools}
\usepackage{graphicx}
\usepackage[table]{xcolor}
\usepackage{chngcntr}
\usepackage{blindtext}
\usepackage{hyperref}
\usepackage{tikz}
\def\BibTeX{{\rm B\kern-.05em{\sc i\kern-.025em b}\kern-.08em
    T\kern-.1667em\lower.7ex\hbox{E}\kern-.125emX}}
    
\usepackage{optidef}
\usepackage{accents}
\usepackage{dsfont}
\usepackage{siunitx}
\usepackage{overpic}
\usepackage{amsmath,amssymb,amsfonts}
\usepackage{graphicx}
\usepackage{textcomp}
\usepackage{xcolor}
\usepackage{optidef}

\usepackage{subcaption}
\usepackage{caption}

\usepackage{lipsum}

\usepackage{amsthm}
\usepackage{array}
\usepackage{color}
\usepackage{url}
 
\usepackage{optidef}
\usepackage{times}
\usepackage{fancyhdr,graphicx,amsmath,amssymb}
\usepackage[ruled,linesnumbered,noend]{algorithm2e}

\usepackage{algorithmicx}
\usepackage{algcompatible}

\IEEEoverridecommandlockouts
\usepackage{cite}
\usepackage{graphicx}
\usepackage{url}
\usepackage{relsize}
\usepackage{hhline}
\usepackage{nicefrac}
\usepackage{amssymb,bm,upgreek}
\usepackage[table]{xcolor}
\usepackage{multirow}

\begin{document}
\title{When Every Bit Counts in Quantized Beamforming: From CSI Feedback to Low-Resolution Hardware}

\author{\IEEEauthorblockN{\normalsize 
    Yasaman Khorsandmanesh,  Parisa Ramezani, and Emil Björnson }
 \thanks{  Y. Khorsandmanesh is with KU Leuven, Leuven, Belgium; P. Ramezani is with University of Technology Sydney, Sydney, Australia; E. Bj\"ornson is with KTH Royal Institute of Technology, Stockholm, Sweden.
}
\thanks{This work was supported by the Grant 2022-04222 from the Swedish Research Council.}
}

\maketitle

\begin{abstract}
Multiple-input multiple-output (MIMO) is a key technology in modern wireless networks, but its performance is often analyzed under ideal hardware assumptions. In practical implementations, however, quantization appears throughout the signal-processing chain, including channel state information (CSI) feedback, fronthaul signaling, beamforming coefficients, data converters, and finite-resolution phase shifters. These impairments do not act independently: they alter the channel information available for beamforming, constrain the feasible transmit signals, and can ultimately limit beamforming gain and interference suppression. This article provides a perspective on quantized beamforming by connecting its main sources, performance implications, and mitigation strategies. We discuss codebook-based CSI feedback, quantization-aware beamforming, low-resolution converter modeling, and finite-resolution hybrid beamforming, and use representative numerical examples to illustrate when quantization becomes a dominant performance bottleneck. We further highlight emerging directions in joint beamforming and quantization design, adaptive-resolution architectures,  and new antenna technologies with discrete hardware constraints. The key message is that finite-resolution effects should be modeled explicitly and incorporated into the beamforming design rather than treated as an afterthought.
\end{abstract}
\vspace{-2mm}
\section{Introduction}

Beamforming in multiple-input multiple-output (MIMO) systems is commonly designed assuming arbitrary precision. In practice, however, channel state information (CSI), beamforming coefficients, and transmitted signals must be represented with a finite number of bits. The resulting quantization errors can reduce beamforming gains and weaken interference suppression, particularly in MIMO systems serving multiple user equipments (UEs) \cite{bjornson2017massive}.

Quantized beamforming differs fundamentally from the impairments in radio-frequency hardware encountered in classical wireless systems. Unlike thermal noise or analog distortions, quantization effects depend explicitly on the signal statistics, the bit resolution, and the architectural split between centralized and distributed processing units \cite{fan2015uplink}. In practical MIMO systems, the quantization of beamforming matrices is closely tied to CSI quality, beamforming design, and resource allocation across UEs, making isolated treatment of these components inadequate. In the digital domain, the main sources of impairments that can be modeled as finite-resolution quantization include limited-rate CSI feedback in frequency-division duplex (FDD) systems \cite{jindal2006mimo}, capacity-constrained fronthaul links \cite{femenias2019cell,khorsandmanesh2023fronthaul}, and finite-resolution analog-to-digital and digital-to-analog converters (ADC) and (DAC) \cite{jacobsson2017quantized,jacobsson2019linear}. In the analog domain, quantization arises from finite-resolution phase shifters (PSs) in hybrid beamforming architectures, particularly at cmWave and mmWave frequencies \cite{sohrabi2016hybrid}. Although these constraints originate from different parts of the system, they all affect the information or signals used for spatial processing and can influence beamforming gain and interference suppression.

This article provides a system-level perspective on quantized beamforming under these impairments. We review their main sources, illustrate their impact through representative numerical examples, discuss quantization-aware mitigation strategies, and highlight emerging research directions. We examine their impact on performance metrics, including the normalized mean squared error (NMSE) and average sum rate. The key design principle is that quantization constraints should be incorporated into beamforming optimization rather than considered only after the beamforming has been designed.


\vspace{-1mm}
\section{Sources of Quantization}
\vspace{-1mm}
Quantization occurs at different stages of a wireless transceiver, introducing trade-offs between communication performance and implementation complexity. We distinguish between quantization on digital processing chain and finite-resolution constraints in the analog front end.
\subsection{Quantization in the Digital Signal Chain}
\begin{figure}
    \centering
    \includegraphics[width=\linewidth]{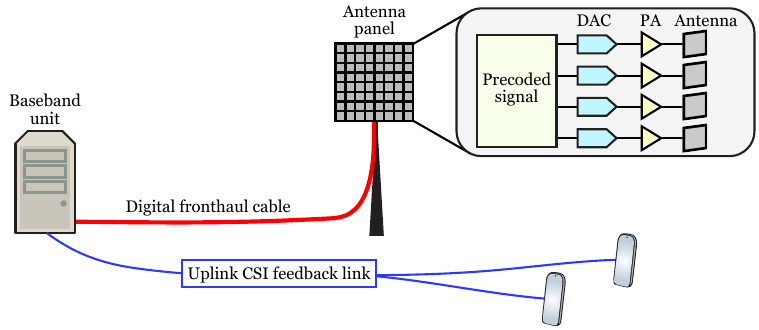}
    \caption{Main sources of quantization in a digital MIMO system.}
    \label{fig:sources}
    \vspace{-4mm}
\end{figure}
As illustrated in Fig.~\ref{fig:sources}, quantization in digital communication systems can appear in several components of the processing chain. In particular, three main sources are commonly encountered in practice: limited CSI feedback, limited fronthaul capacity, and the use of DACs and ADCs with limited resolution which has direct effect on digital beamforming.

\subsubsection{Limited Feedback CSI}
The base station (BS) requires CSI to design efficient beamforming that directs the transmitted energy toward the intended UEs and suppresses interference to other UEs. However, in many practical systems, especially in FDD mode, the BS cannot directly estimate the downlink channel. Instead, the UE estimates it and sends this information back to the BS through an uplink feedback link. This feedback channel has limited capacity and strict latency requirements. Transmitting full-resolution CSI would require a large number of bits, particularly in massive MIMO systems, since the channel matrix dimension grows with the number of antennas. Therefore, sending the complete CSI is often not feasible. In such systems, the UE quantizes the estimated channel before sending it to the BS. The BS designs the downlink beamforming based on the quantized CSI received from the UEs. This quantization inevitably introduces inaccuracies in the channel representation. As a result, the computed beamforming vectors may deviate from the optimal ones, thereby limiting the beamforming’s ability. 
\subsubsection{Fronthaul Capacity Constraints}
The architecture of 5G BSs introduces a new system bottleneck: the digital fronthaul link between the antenna panel (AP) and the baseband unit (BBU). The fronthaul is responsible for transporting both channel estimates and precoded data streams. The required fronthaul rate increases with the number of AP antennas and the signal sampling rate (i.e., bandwidth). Since the fronthaul capacity is limited in practical deployments, the exchanged signals must be quantized before transmission. This quantization reduces signal resolution and introduces distortion, affecting the performance of both downlink beamforming and uplink channel estimation in time-division duplex (TDD) systems. Similar limitations appear in \emph{cell-free massive MIMO} systems, where many distributed access points are connected to a central processing unit through fronthaul links with limited capacity. In such systems, efficient quantization methods are important to maintain scalability and communication performance.
\subsubsection{Low-Resolution DACs/ADCs}
Another important source of quantization that effects digital beamforming appears in DACs and ADCs in transceivers. These components convert digital baseband signals into analog waveforms and vice versa. The continuous-valued digital baseband samples are first represented using a finite set of amplitude levels before being converted into analog signals. If the DAC resolution is low, the output waveform is only a rough approximation of the intended signal, which introduces distortion before transmission. Similarly,  the ADC maps the received analog waveform to a finite number of digital levels, thereby losing some amplitude and phase information during sampling and quantization. This loss becomes more significant when only a few quantization bits are used, and it can degrade channel estimation, detection, and beamforming performance.

In systems with high-resolution DACs/ADCs, the converted signal closely approximates its ideal unquantized representation, but such hardware is power-hungry and requires complex circuitry. This becomes particularly challenging in wideband mmWave systems and massive MIMO deployments where hundreds of antennas may be used. For this reason, low-resolution DACs and ADCs are used to reduce hardware costs and power consumption.

\subsection{Finite-Resolution Analog Beamforming}
PSs are a key component in analog and hybrid beamforming transceivers. In ideal models, the PSs can realize any continuous phase value. In practice, however, hardware constraints allow only discrete phase values. The desired phase shifts must therefore be quantized before they are applied. In analog beamforming, the signal direction is usually controlled by these PSs. Phase quantization creates a mismatch between the intended and implemented beamforming matrix. Consequently, the beam may not be steered exactly to the desired direction, thereby reducing beamforming gain and increasing interference. To address this limitation, several analog and hybrid transceiver designs explicitly account for discrete PSs. When the PS's resolution is moderate, jointly designing the analog and digital beamforming matrices allows hybrid beamforming systems with finite-resolution PSs to achieve performance close to that of ideal high-resolution architectures. Nevertheless, discrete phase shifts remain an important practical source of quantization error in analog and hybrid beamforming systems \cite{rajashekar2016hybrid}.

\subsection{Performance Impact of Different Quantization Sources}

This subsection investigates how quantization at different stages of the transmission chain affects beamforming performance in a downlink MIMO system. We compare the ideal unquantized Wiener-filter (WF) beamforming \cite{joham2005linear}, also known as regularized zero-forcing, with systems subject to CSI quantization, limited fronthaul capacity, low-resolution DACs, and finite-resolution PSs. We use uniform scalar quantization as in \cite{hui2001unifquantized} where applicable, a DFT codebook for CSI quantization, and a uniformly spaced discrete phase set for finite-resolution PSs.

The purpose is to illustrate how these impairments affect the NMSE performance individually and cumulatively under different channel conditions. Note that all curves are evaluated using a hybrid beamforming architecture with mmWave channels, where the beamforming matrix consists of a digital beamforming followed by an analog beamforming.

The results are obtained using Monte Carlo simulations with a Rician fading model. In each channel realization, the propagation channel contains both a deterministic line-of-sight component and a scattered non-line-of-sight component, where the relative strength is controlled by the Rician factor $\kappa$. All UEs are assumed to have the same SNR, so the observed performance differences are mainly due to the considered quantization sources rather than large-scale fading variations among UEs. We consider QPSK modulation. The curves are based on the following schemes:
\begin{itemize}
\item Inf. Res., which uses WF beamforming without any quantization;
\item QCSI, which uses the discrete Fourier transform (DFT) codebook for quantized CSI and where the digital and analog beamforming matrices are designed based on that;
\item QBF, where the digital beamforming coefficients are quantized;
\item QDAC; in which the digitally precoded transmit signals are quantized;
\item QPS; where the analog beamforming matrix is constrained to have quantized phase values rather than continuous ones;
\end{itemize}

\begin{figure}
    \centering
    \includegraphics[width=0.7\linewidth]{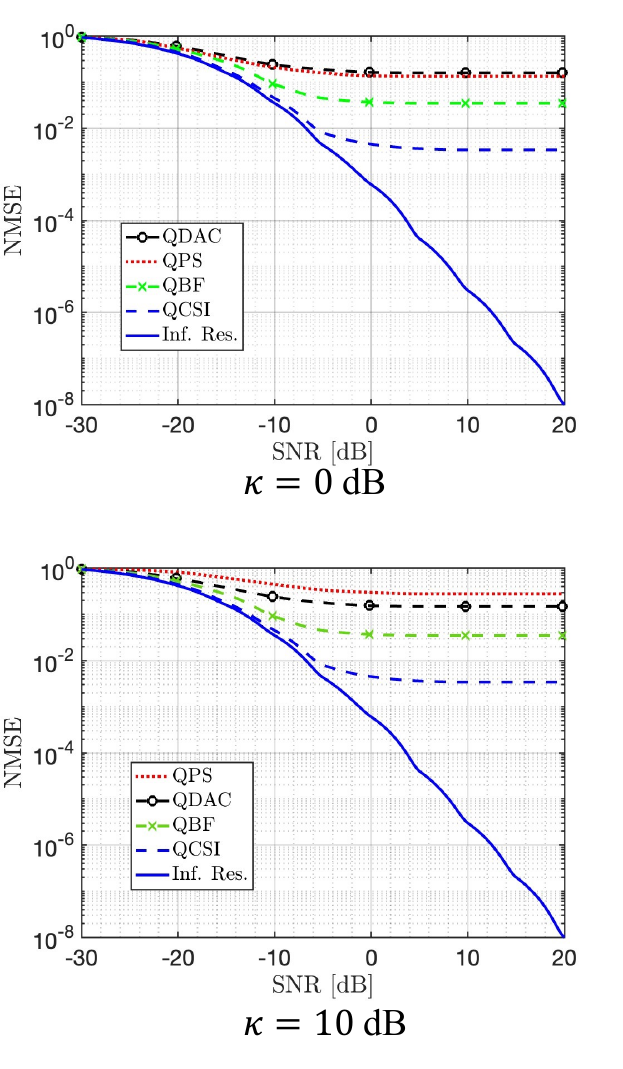} 
    \caption{NMSE versus SNR for MIMO system with $M=8$, $K=1$, and $L=2$.}
\label{fig:quantization_sources}
\end{figure}

Fig.~\ref{fig:quantization_sources}  shows the NMSE performance considering different quantization sources where AP contains $M=8$  antennas, serving  $K=1$ single-antenna UE with $L=2$ levels of quantization (1 bit resolution per real domain) for two extreme values of $\kappa$. In both cases, Inf. Res. gives the lowest NMSE and decreases rapidly with SNR. The quantized cases, however, reach error floors at high SNR, which shows that quantization becomes the dominant limitation once the noise level is sufficiently small. The comparison between $\kappa=10$ dB and $\kappa=0$ dB shows that the relative impact of the quantization sources depends on the channel conditions. For larger $\kappa$, the channel is dominated by a strong line-of-sight component, so CSI and phase quantization errors can cause noticeable beam misalignment and beamforming-gain loss. For smaller $\kappa$, the channel has richer scattering, and the transmitted signal depends on a more complex spatial structure. In this case, 'QDAC' can have a stronger impact because it directly distorts the spatial signal representation.

\begin{figure}
    \centering
    \includegraphics[width=0.7\linewidth]{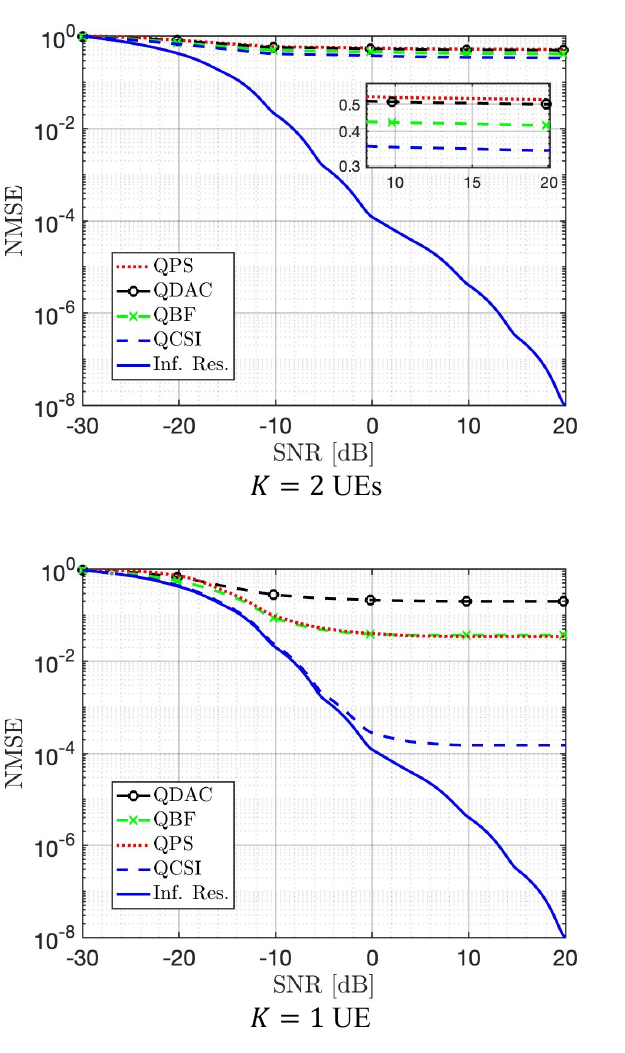}
    \caption{NMSE versus SNR for MIMO system with $M=16$,  $L=4$, and $\kappa=4$ dB.}
    \label{fig:differentusers}
\end{figure}

Fig.~\ref{fig:differentusers} considers a larger  system with $M=16$ and $L=4$ and comparing $K=1$ and  $K=2$ UE setups, with $\kappa=4$ dB. For $K=2$  system, the quantized curves have higher error floors. This is expected since multi-user transmission requires accurate interference suppression. For this system, 'QBF' and 'QPS' almost have similar effects. Since several parameters differ from the previous figures, including $M$, $L$, and $\kappa$, this result should not be interpreted as isolating the impact of increasing the number of antennas. The results instead show that the changed spatial configuration can improve performance in both ideal and some quantized cases. Increasing the number of UEs generally increases the NMSE across the considered quantization sources because multi-user transmission places stricter requirements on interference suppression. This result further highlights the importance of accounting for quantization in multi-user system design

\begin{figure}[t]
    \centering
    \includegraphics[width=0.7\linewidth]{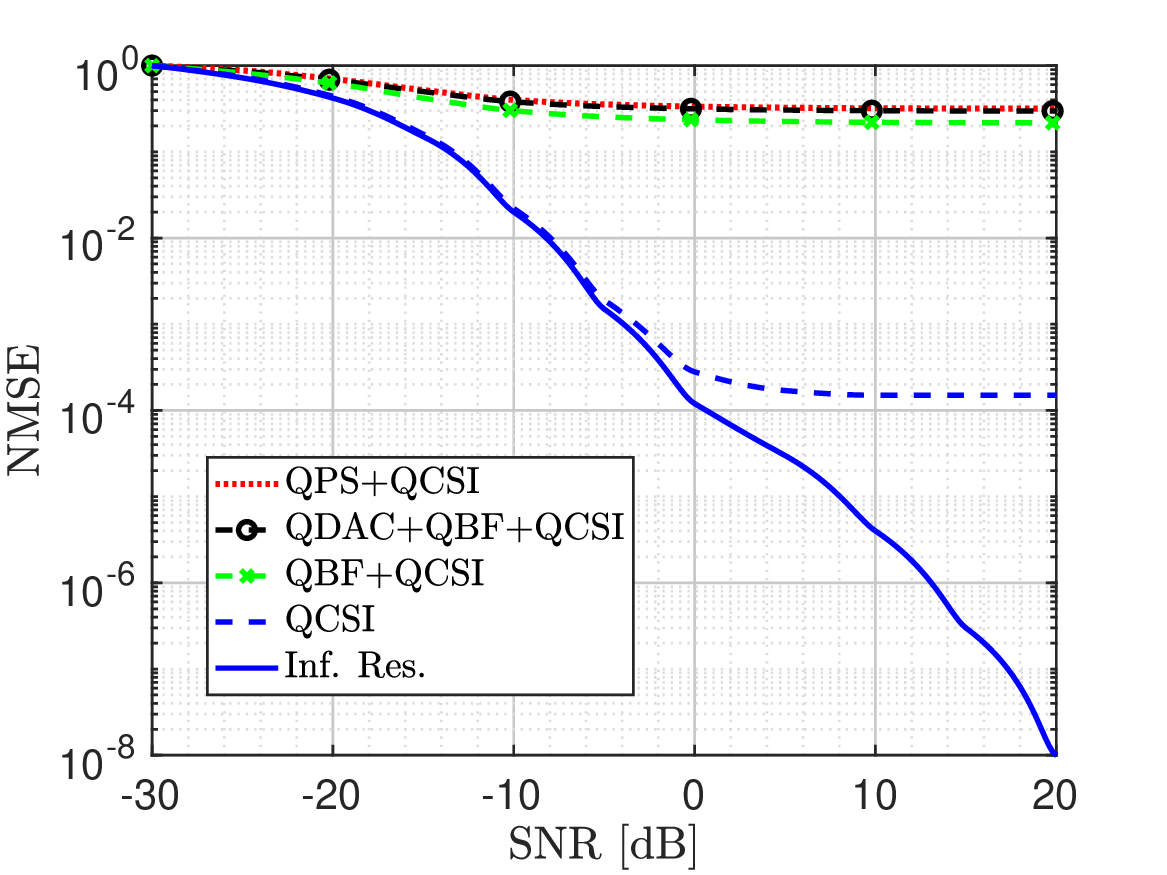}
    \caption{NMSE versus SNR for cumulative quantization impairments for MIMO system with $M=16$, $K=1$, $L=4$, and $\kappa=4$ dB.}
    \label{fig:cumulative}
\end{figure}

Fig.~\ref{fig:cumulative} illustrates cumulative quantization impairments with the same setups as Fig.~\ref{fig:quantization_sources}. As expected, each additional quantization stage increases the NMSE floor, confirming that practical quantization sources should be studied together, since their effects accumulate in the transmitted signal. For the analog architecture, the considered case includes limited CSI feedback and finite-resolution PSs. The analog beamforming case exhibits higher NMSE than the digital architecture, even when all digital quantization sources are included. This degradation is mainly due to the limited flexibility of analog beamforming, which controls only the phases of the transmitted signals, while the amplitudes remain unchanged.

Overall, the results show that quantization has a limited impact at low SNR, where noise dominates, but it becomes the main bottleneck at high SNR. In the considered scenarios, CSI quantization is often among the dominant impairments, although the relative importance of the different quantization sources depends strongly on the propagation conditions and system configuration. Therefore, practical massive MIMO systems require a proper codebook, quantization-aware beamforming, and hardware-aware design to balance performance, complexity, and implementation cost.

\section{Quantization Challenges and Design Solutions}
\begin{figure*}[t]
  \centering  \includegraphics[width=0.8\linewidth]{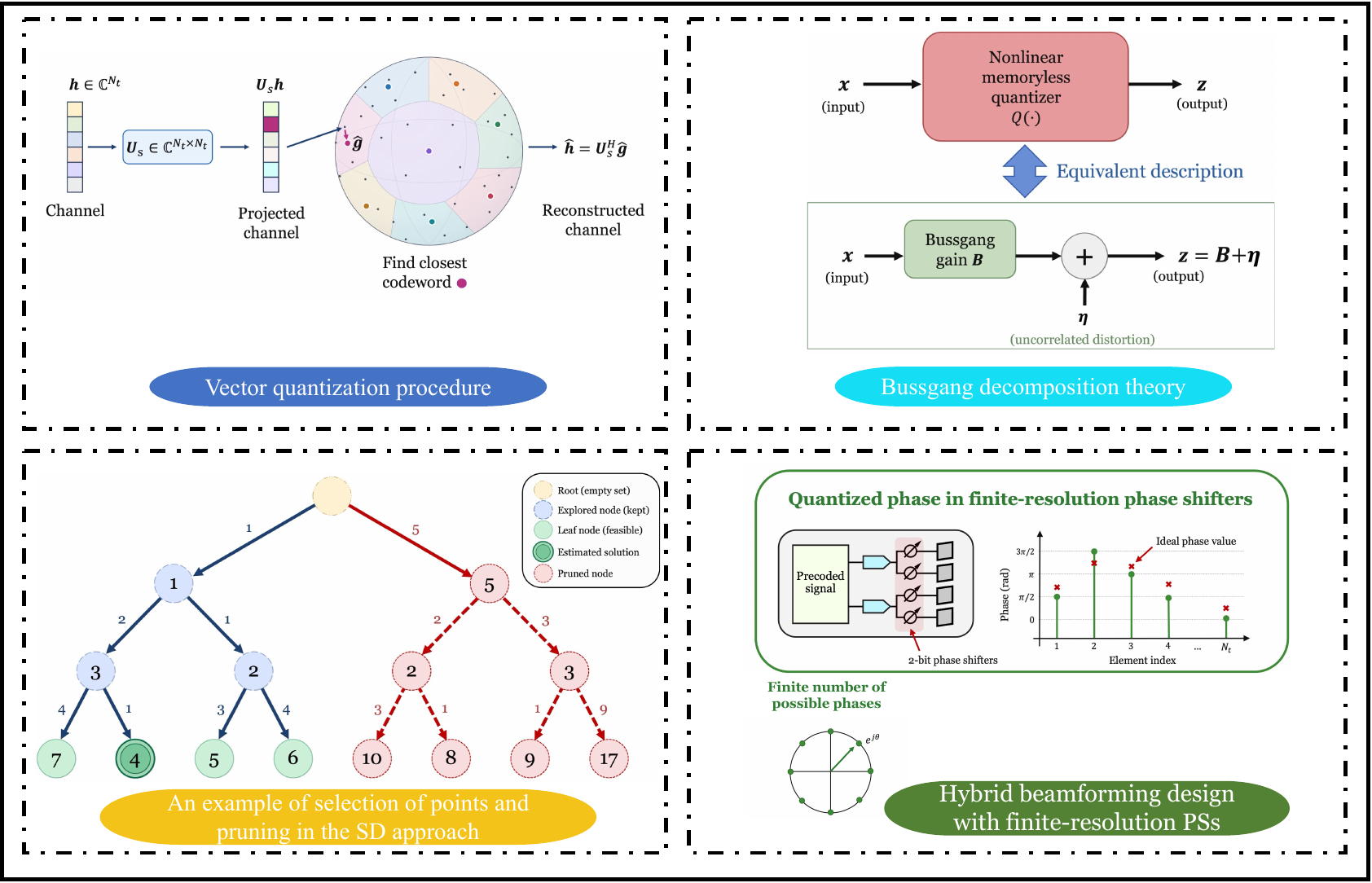}
        \caption{Key methods to address sources of quantization in beamforming design systems.}
        \label{fig:Challenges}

\end{figure*}

Each source of quantization creates a different design challenge. Limited CSI feedback mainly leads to an inaccurate channel representation, quantization in fronthaul and converters introduce distortion in the beamforming matrix or transmitted signal, and finite-resolution PSs restrict the set of feasible analog beamforming directions. Fig. \ref{fig:Challenges} illustrates representative mitigation strategies for the main sources of quantization, which we explain in the following subsections.

\subsection{Limited Feedback CSI}
Limited-feedback CSI is commonly handled using codebooks. A codebook is a predefined set of candidate channels known by both the BBU and the UE. Instead of separately feeding back each entry of the UE's channel estimate, the UE selects the codebook entry that best matches its current estimate and sends only the corresponding index to the BBU. This approach can vastly reduce the required feedback load, but the quantization accuracy depends strongly on the codebook design.

A good codebook should represent the ``likely'' spatial channel directions with sufficient accuracy while using a limited number of feedback bits. Hence, the appropriate codebook structure depends on the propagation environment. For channels with approximately isotropic fading, all directions in the vector space are almost equally likely. Grassmannian codebooks are the preferred choice in this case \cite{el2012grassmannian}. 
By contrast, channels with a dominant angular component (e.g., line-of-sight path) are better represented by array-response-based codebooks.
A DFT-based codebook with an appropriate oversampling factor is the appropriate choice when using uniform linear arrays, since their array response vectors are DFT vectors.
Generally speaking, the principle of efficient codebook design is to transform the channel to a suitable vector space (possibly of lower dimension), create a quantization grid in that space, and use the nearest-neighboring rule when selecting what grid point to feed back. Modern codebook designs, therefore, exploit channel correlation across space, frequency, and time to reduce the feedback overhead while preserving the dominant channel structure. This principle is also central to current 3GPP CSI feedback mechanisms. However, fixed codebooks may become inefficient when the channel statistics change or the channel changes rapidly. A promising direction for 6G is therefore to complement structured codebooks with learned parameters or adaptive compression mappings, while retaining a standardized and interpretable feedback procedure.

\subsection{Fronthaul Capacity Constraints}
When fronthaul capacity is limited, the system must carefully decide which information to exchange between the BBU and the AP. One possibility is to compute the entire beamforming matrix centrally and then quantize it before sending it to the AP. Another possibility is to split the processing so that part of the beamforming is performed closer to the antennas at the AP. In both cases, the main challenge is that fronthaul quantization changes the spatial structure of the transmitted signal. As a result, quantization errors in the beamforming matrix may reduce the desired signal power and make multi-user interference cancellation less effective \cite{femenias2019cell}.

In contrast to conventional C-RAN settings, the large number of antennas in massive MIMO systems makes beamformed signaling a dominant part of the fronthaul load. This motivates separating data and beamforming matrix transmission over the fronthaul. The BBU sends the user data to the AP, where it is mapped to modulation symbols, while the precoding matrix is quantized and transmitted separately. The AP then applies the quantized beamforming matrix to the data symbols before wireless transmission. Our numerical results show that directly quantizing the beamformed signal achieves slightly lower MSE at the same resolution. However, separate signaling requires substantially less fronthaul capacity and therefore provides a more favorable performance–fronthaul trade-off for large-scale massive MIMO systems.

If the beamforming matrix is designed first under ideal assumptions and then quantized, the resulting system can be far from optimal \cite{khorsandmanesh2023fronthaul}.
Hence, the key design principle is to treat quantization as an integral part of the beamforming matrix design.
By including the fronthaul constraint directly in the beamforming matrix design, one can obtain integer program formulations and solve them using techniques resembling those originally designed for MIMO detection. Sphere decoding is an example of such a technique that provides an efficient solution by searching only for candidate transmit vectors that lie within a sphere around the desired received signal. Branches that cannot lead to a better solution are removed early, reducing search complexity compared with exhaustive enumeration while preserving near-optimal performance for such systems. Fig.~\ref{fig:Challenges} illustrates how sphere decoding selects promising candidate points while pruning branches that cannot improve the solution.

 \subsubsection*{\textbf{Results}}
\begin{figure}[t]
    \centering
    \includegraphics[width=0.7\linewidth]{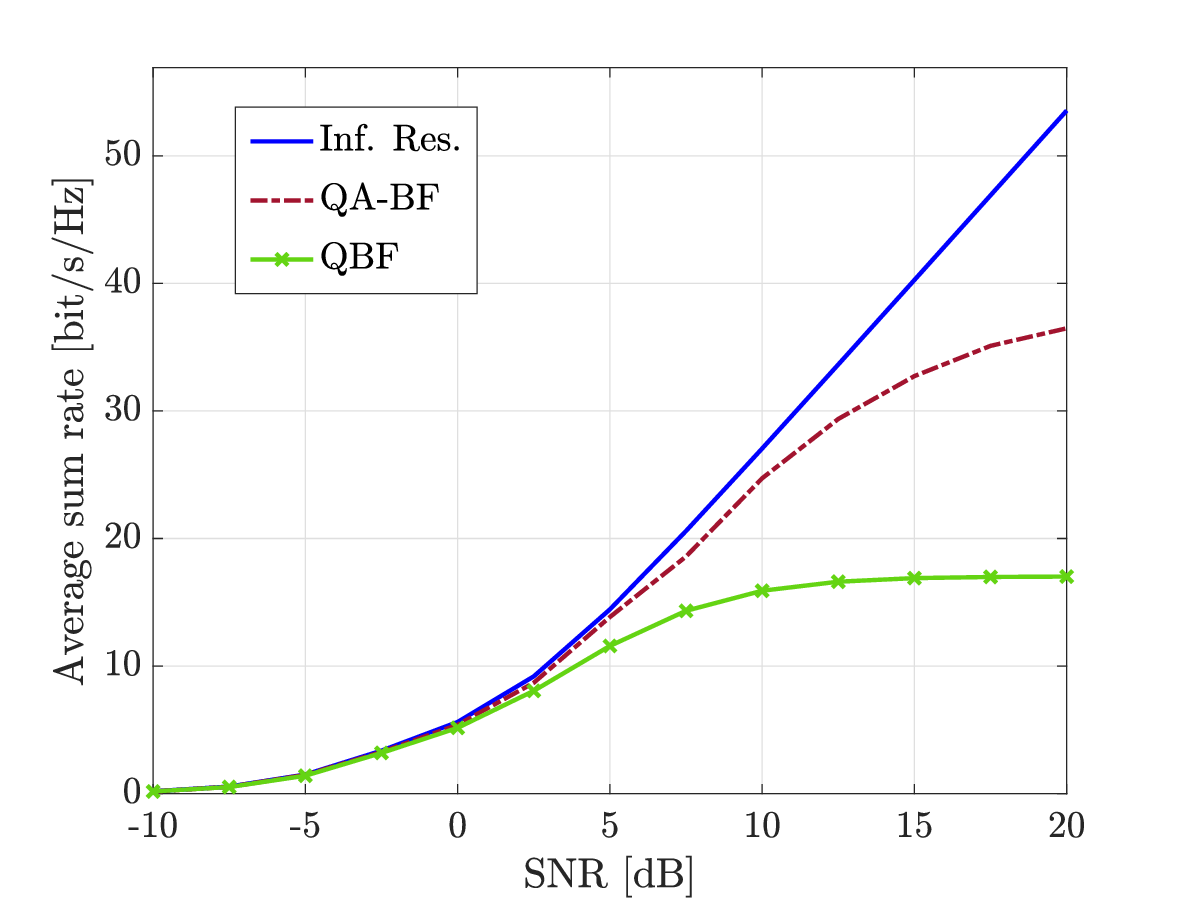}
    \caption{Comparing the different effects of integrating quantization for a fronthaul limited scenario, assuming $M=16$, $K=4$, and $L=8$.}
    \label{fig:challenge}
\end{figure}

To illustrate the benefit of quantization-aware design, Fig.~\ref{fig:challenge} shows the average sum rate versus SNR for different precoding schemes. The setup is $M = 16$, $L=8$, and $K = 4$. The \emph{Inf. Res.} curve represents ideal precoding without quantization and provides the highest rate. In contrast, all finite-resolution schemes saturate at high SNR because quantization prevents complete interference cancellation, making the system interference-limited. The proposed sphere-decoding-based quantization-aware beamforming (\emph{QA-BF}) substantially reduces the gap to the infinite-resolution benchmark compared with conventional quantized beamforming (\emph{QBF}), where the ideal beamforming matrix is designed first and then quantized. At high SNR, \emph{QA-BF} achieves approximately twice the sum rate of \emph{QBF}, demonstrating the advantage of incorporating the quantization constraints directly into the beamforming matrix design.

\subsection{Low-Resolution DACs and ADCs}
Low-resolution DACs and ADCs introduce nonlinear distortion that depends on the transmitted or received signal. When the resolution is moderate, linear beamforming followed by quantization can still perform well, especially in massive MIMO systems where the large number of antennas provides an averaging effect that suppress distortion \cite{jacobsson2019linear}.

However, when the resolution is very low, such as in one-bit DAC systems, conventional linear beamforming performs poorly. The transmitted signal can only take a small number of possible values, so the system cannot accurately represent the continuous-valued signal produced by a standard beamforming matrix. In this case, nonlinear beamforming becomes important. Instead of designing a continuous-valued signal and then quantizing it, nonlinear methods directly search for a transmit signal from the finite alphabet supported by the hardware. This approach can better control the distortion and reduce inter-user interference, but it also leads to more difficult optimization problems. Several approximate methods have been proposed to make these problems tractable, including convex relaxation, semidefinite relaxation, iterative projection methods, message-passing algorithms, and tree-search techniques. Such methods are more complex than linear beamforming, but they can provide large gains when the resolution is very low.

A common way to handle quantization analytically is to use an equivalent linear model. In particular, for Gaussian or approximately Gaussian quantizer inputs, the Bussgang decomposition represents the output of a finite-resolution quantizer as a scaled version of the input plus an uncorrelated distortion term, as depicted in Fig.~\ref{fig:Challenges}. This does not make the quantizer linear, but it preserves useful second-order statistics and enables the application of standard signal-processing tools. The resulting quantized transmitted signal design can then target objectives such as sum-rate maximization, mean-squared-error minimization, or energy-efficient operation under fronthaul and hardware constraints \cite{jacobsson2017quantized}. Since the distortion depends on the signal statistics, these problems are generally nonconvex and are often solved using iterative methods such as alternating optimization, successive convex approximation, or weighted mean-square-error formulations.

\subsection{Finite-Resolution PSs}
A simple approach is to first design an ideal continuous-phase beamformer and then quantize each phase to the nearest available value. This method is easy to implement, but it treats each phase independently and ignores the coupling between antenna elements. Fig.~\ref{fig:Challenges} shows this method.

Such independent quantization can accumulate errors across the array. The resulting beam may have lower gain in the desired direction or higher sidelobes in unwanted directions. The degradation is particularly visible when the phase resolution is very low or when the system operates in an interference-limited regime. More advanced methods include the discrete nature of the PSs directly in the beamformer design. In hybrid beamforming, this often means jointly optimizing the analog beamforming and the digital baseband beamforming matrix, while accounting for the limited resolution of analog PSs in the optimization process. Iterative algorithms are commonly used because such problems are difficult to solve globally. These algorithms alternate between updating the analog and digital parts while respecting the finite-resolution constraints \cite{sohrabi2016hybrid}. In the extreme case of one-bit PSs, each phase can take only two values. While an exhaustive search may be a viable option for small antenna arrays, it becomes infeasible for large antenna arrays. Approximate methods based on rank approximations, surrogate objectives, or low-complexity iterative updates can provide good performance with manageable complexity.

\section{Emerging Research Directions}
The methods discussed above demonstrate how incorporating finite-resolution constraints into beamforming design can mitigate the detrimental impact of quantization at different stages of the communication chain. Their interaction, however, raises broader system-design questions. The quantization errors depend on the signals and hardware settings, so that these errors can be combined across antennas and affect the received signals. This dependence creates opportunities to jointly design quantization, beamforming, and antenna configurations according to application requirements and energy constraints. In the following, we outline emerging research directions that build on this perspective.

\subsection{Joint Design of Quantization and Beamforming}

In many systems, beamforming and quantization are still treated as separate problems. First, a beamforming matrix is designed assuming ideal hardware. Then, the resulting coefficients or signals are quantized. This sequential approach is simple but often suboptimal.  A more promising direction is to jointly optimize beamforming and quantization while accounting for hardware and architectural constraints. Depending on the system, this may involve beamforming coefficients, CSI feedback, fronthaul bit allocation, converter resolution, or PSs states. The resulting problems are typically nonconvex and involve both continuous and discrete variables. Model-based approaches, such as alternating optimization, successive approximation, and mixed-integer optimization, can address these challenges. Learning-based methods can complement them by adapting quantization policies, bit allocation, and distortion compensation to changing channel and hardware conditions. However, ensuring robustness, interpretability, and computational efficiency remains challenging, motivating hybrid model- and learning-based designs.


\subsection{Multi-Objective Quantization}

Most quantization methods are designed to minimize a signal-reconstruction metric, such as the mean-squared error between the original and quantized signals. However, accurate reconstruction of all signal components may not be necessary when the quantized information is only used for a specific downstream task. In such cases, the quantizer can instead be designed to preserve the information most relevant to the intended application. This approach, often referred to as task-specific quantization, can be applied to communication, sensing, localization, detection, or control, where the quantization strategy is optimized for the corresponding task performance rather than waveform reconstruction. An interesting research direction is therefore to jointly design beamforming and quantization according to task-level objectives that go beyond maximization of the signal-to-noise-and-distortion ratio (SNDR). 

An additional opportunity arises when the quantization resolution can be adjusted after deployment. This flexibility is particularly relevant to digital quantization, where the number of bits can be adapted according to channel conditions and system requirements. In contrast, adapting the resolution of DACs, ADCs, and PSs requires hardware that supports such reconfiguration. Also, sustainability and energy efficiency are becoming increasingly important in future wireless networks, with 6G placing particular emphasis on these aspects compared with previous generations \cite{zeadally2012energy}. This trend motivates multi-objective quantization, where the goal is not only to achieve satisfactory task performance but also to reduce energy consumption, signaling overhead, or implementation complexity. Jointly optimizing the quantization resolution and beamforming strategy could therefore provide a promising way to balance these competing objectives, enabling more energy-efficient and adaptive MIMO systems.


\subsection{Quantized beamforming with Movable Antennas}

Movable antenna systems have recently emerged as a new architecture for improving wireless performance without requiring more antennas by allowing the physical positions of antenna elements to be adjusted within a given region \cite{zhu2023movable}. This architecture fundamentally changes how quantized beamforming problems should be approached. In particular, because spatial steering vectors vary continuously with antenna movement, conventional fixed-geometry codebooks for limited CSI feedback may prove insufficient, motivating position-aware codebook designs. Moreover, existing movable antenna designs can differ in how the antenna locations are modeled: some optimize the antenna positions over a continuous spatial region, while others restrict them to a finite set of candidate positions on a discrete grid. The latter can be interpreted as a form of location-based quantization, where continuous antenna positions are represented by a finite set of feasible spatial points. This location quantization is fundamentally coupled to beamforming optimization because changing antenna locations reshapes the channel matrix, which in turn makes quantization distortion and multi-user interference position-dependent. These new challenges motivate the development of algorithms that jointly optimize discrete antenna locations, position-aware codebooks, and quantized beamforming matrices, while accounting for both location quantization errors and position-dependent quantization effects.

\subsection{Quantized Tri-Hybrid Beamforming with Reconfigurable Antennas}

Another emerging direction is tri-hybrid MIMO, in which conventional digital and analog beamforming are complemented by a third electromagnetic beamforming layer implemented with reconfigurable antennas \cite{deshpande2026signal}. This architecture offers design flexibility similar to that of fully digital beamforming while introducing new finite-resolution design challenges. In particular, the analog beamforming stage may rely on low-resolution PSs, while the electromagnetic properties or configurations of the reconfigurable antennas may also be selected from discrete sets. These constraints motivate the joint design of digital precoding, quantized phase shifts, and discrete antenna configurations. Unlike conventional hybrid beamforming, the quantization errors introduced at the analog and electromagnetic layers are coupled, since changing the antenna configuration also modifies the effective channel seen by the digital and analog beamforming. Future work should therefore investigate quantization-aware tri-hybrid beamforming algorithms that jointly determine the digital beamforming, finite-resolution PS settings, and reconfigurable antenna states while balancing spectral efficiency, energy consumption, and implementation complexity.

\section{Conclusion}
This article has highlighted quantization as a system-level constraint in practical massive MIMO, arising in CSI feedback, fronthaul transmission, ADCs, DACs, and PSs. Its impact is operating-point dependent: it is often minor at low SNR but becomes performance-limiting at high SNR and in multi-user systems, where accurate interference suppression is essential.
Codebooks, Bussgang-based models, and hardware-aware beamforming offer complementary mitigation strategies. Quantized beamforming remains an open research area, particularly for joint beamforming and bit allocation, learning-assisted compensation, movable antennas, and energy-efficient architectures.


\bibliographystyle{IEEEtran}
\bibliography{references}

\vspace{-11mm}
\begin{IEEEbiographynophoto}{Yasaman Khorsandmanesh} received the
B.Sc. degree in telecommunication (electrical engineering) from the
K. N. Toosi University of Technology, Iran, in 2018, M.Sc. and Ph.D. degrees in
communication systems from the Amirkabir University of Technology, Iran, in 2020, and from the
KTH Royal Institute of Technology, Sweden, in 2026, respectively. She is currently a postdoctoral researcher at KU Leuven, Belgium. Her research focuses on signal processing and wireless communications.
\end{IEEEbiographynophoto}
\vspace{-9mm}
\begin{IEEEbiographynophoto}{Parisa Ramezani}
received the
B.Sc. degree in electrical engineering from the
Sharif University of Technology, Iran,
in 2014, and the M.Phil. and Ph.D. degrees in
electrical and information engineering from the
University of Sydney, Australia, in 2017
and 2022, respectively. From 2022 to 2026, she was a Postdoctoral Researcher with the KTH Royal
Institute of Technology, Sweden. She is currently a chancellor's research fellow at the University of Technology Sydney, Australia. Her
research focuses on wireless communications and
signal processing.
\end{IEEEbiographynophoto}
\vspace{-11mm}
\begin{IEEEbiographynophoto}{Emil Bj\"ornson} 
 received the M.S. degree in engineering mathematics from
Lund University, Sweden, in 2007, and the Ph.D.
degree in telecommunications from the KTH
Royal Institute of Technology, Sweden, in 2011.
He is a Professor of Wireless Communication
with the KTH Royal Institute of Technology,
Stockholm, Sweden. He has a podcast and YouTube channel called Wireless Future, has
authored four textbooks, and published much
simulation code. His research focuses on multiantenna communications and radio resource management, using methods
from communication theory, signal processing, and machine learning. He
is a Digital Futures Fellow, a Wallenberg Academy Fellow, and a Clarivate
Highly Cited Researcher.
\end{IEEEbiographynophoto}
\end{document}